\documentstyle[aas2pp4,epsf,astrobib]{article}
\def\lastrev{27 Nov 2000}

\let\oldfootsep=\footnotesep
\def\simlt{\hbox{ \rlap{\raise 0.425ex\hbox{$<$}}\lower 0.65ex\hbox{$\sim$} }}
\def\simgt{\hbox{ \rlap{\raise 0.425ex\hbox{$>$}}\lower 0.65ex\hbox{$\sim$} }}

\def\that{{\hat t}}

\def\msun{M_\odot}

\def\that{{\widehat t}\,} 

\def\leaderfill{\leaders\hbox to 1em{\hss-\hss}\hfill}

\def\etal{{\rm et al.\ }}

\def\kms{{\rm\, km/s}}
\def\kpc{{\rm\, kpc}}
\def\pc{{\rm\, pc}}

\def\ten#1{\times 10^{#1}}
\def\msun {{ \rm \, M_\odot}}

\def\lsim{\mathrel{\mathpalette\@versim<}}
\def\gsim{\mathrel{\mathpalette\@versim>}}
\def\spose#1{\hbox to 0pt{#1\hss}}
\def\simlt{\mathrel{\spose{\lower 3pt\hbox{$\mathchar"218$}}
     \raise 2.0pt\hbox{$\mathchar"13C$}}}
\def\simgt{\mathrel{\spose{\lower 3pt\hbox{$\mathchar"218$}}
     \raise 2.0pt\hbox{$\mathchar"13E$}}}

\begin{document}

\title{MACHO Project Limits on Black Hole Dark Matter 
in the 1-30 Solar Mass Range.}

\author{
      C.~Alcock\altaffilmark{1,2},
    R.A.~Allsman\altaffilmark{3},
      D.R.~Alves\altaffilmark{12},
    T.S.~Axelrod\altaffilmark{4},
      A.C.~Becker\altaffilmark{6},
    D.P.~Bennett\altaffilmark{10,1},
    K.H.~Cook\altaffilmark{1,2},
      N.~Dalal\altaffilmark{2,5},
    A.J.~Drake\altaffilmark{1,4},
    K.C.~Freeman\altaffilmark{4},
      M.~Geha\altaffilmark{1},
      K.~Griest\altaffilmark{2,5},
    M.J.~Lehner\altaffilmark{11},
    S.L.~Marshall\altaffilmark{1,2},
    D.~Minniti\altaffilmark{1,13},
    C.A.~Nelson\altaffilmark{1,15},
    B.A.~Peterson\altaffilmark{4},
      P.~Popowski\altaffilmark{1},
    M.R.~Pratt\altaffilmark{6},
    P.J.~Quinn\altaffilmark{14},
    C.W.~Stubbs\altaffilmark{2,4,6,9},
      W.~Sutherland\altaffilmark{7},
    A.B.~Tomaney\altaffilmark{6},
      T.~Vandehei\altaffilmark{2,5},
      D.~L.~Welch\altaffilmark{8}
        }
\begin{center}
{\bf (The MACHO Collaboration) }\\
\lastrev   
\end{center}

\begin{abstract}
\rightskip = 0.0in plus 1em

We report on a search for long duration microlensing events towards the
Large Magellanic Cloud.  We find none, and therefore put limits
on the contribution of high mass objects to the Galactic dark matter.
At 95\% confidence level we exclude objects in
the mass range $0.3 \msun$ to $30.0 \msun$ from
contributing more than $4 \ten{11} \msun$ to the Galactic halo.
Combined with earlier results, this means that objects with masses under
$30 \msun$ cannot make up the entire dark matter halo if the halo is of
typical size.
For a typical dark halo, objects with masses under $10 \msun$ contribute
less than 40\% of the dark matter.

\end{abstract}


\altaffiltext{1}{Lawrence Livermore National Laboratory, Livermore, CA 94550\\
    Email: {\tt alcock, kcook, adrake, mgeha, stuart, dminniti, cnelson,
    popowski@igpp.ucllnl.gov}}

\altaffiltext{2}{Center for Particle Astrophysics,
    University of California, Berkeley, CA 94720}

\altaffiltext{3}{Supercomputing Facility, Australian National University,
    Canberra, ACT 0200, Australia \\
    Email: {\tt Robyn.Allsman@anu.edu.au}}

\altaffiltext{4}{Research School of Astronomy and Astrophysics,
        Canberra, Weston Creek, ACT 2611, Australia\\
 Email: {\tt tsa, kcf, peterson@mso.anu.edu.au}}

\altaffiltext{5}{Department of Physics, University of California,
    San Diego, CA 92093\\
    Email: {\tt endall@physics.ucsd.edu, kgriest@ucsd.edu, vandehei@astrophys.uc
sd.edu }}

\altaffiltext{6}{Departments of Astronomy and Physics,
    University of Washington, Seattle, WA 98195\\
    Email: {\tt becker, stubbs@astro.washington.edu}}

\altaffiltext{7}{Department of Physics, University of Oxford,
    Oxford OX1 3RH, U.K.\\
    Email: {\tt w.sutherland@physics.ox.ac.uk}}

\altaffiltext{8}{McMaster University, Hamilton, Ontario Canada L8S 4M1\\
    Email: {\tt welch@physics.mcmaster.ca}}

\altaffiltext{9}{Visiting Astronomer, Cerro Tololo Inter-American Observatory}

\altaffiltext{10}{Department of Physics, University of Notre Dame, IN 46556\\
    Email: {\tt bennett@bustard.phys.nd.edu}}

\altaffiltext{11}{Department of Physics, University of Sheffield, Sheffield S3 7
RH, UK\\
    Email: {\tt m.lehner@sheffield.ac.uk}}

\altaffiltext{12}{Space Telescope Science Institute, 3700 San Martin Dr.,
    Baltimore, MD 21218\\
    Email: {\tt alves@stsci.edu}}

\altaffiltext{13}{Depto. de Astronomia, P. Universidad Catolica, Casilla 104,
        Santiago 22, Chile\\
Email: {\tt dante@astro.puc.cl}}

\altaffiltext{14}{European Southern Observatory, Karl Schwarzchild Str.\ 2,
        D-8574 8 G\"{a}rching bel M\"{u}nchen, Germany\\
Email: {\tt pjq@eso.org}}

\altaffiltext{15}{Department of Physics, University of California, Berkeley,
        CA 94720}

\setlength{\footnotesep}{\oldfootsep}

\section{Introduction}
\label{sec-intro}

Recent results from the MACHO and EROS collaborations
have ruled out Massive Compact Halo Objects (MACHOs)
as the bulk the Galactic dark matter in the mass range
$10^{-7}\msun$ to a few solar masses \cite{macho-lmc5,eros-2,macho-eros-spike},
thus eliminating the main candidate for baryonic dark matter in the Milky Way.
However, there still remains a window between several solar masses and
around one thousand solar masses \cite{moore}, 
where black holes or other MACHOs, 
could make up the dark matter of the Milky Way.
It was shown by Carr and Hawking (1974) that primordial black
holes could have formed at very early stages in the Universe as
a result of initial inhomogeneities, and recent work has focussed
on the spike in the primordial black hole mass spectrum
that could arise during the quark-hadron
phase transition in the early Universe (e.g. Jedamzik 1997).
As reviewed by Carr (1994), the density relative to critical density,
$\Omega$, in compact dark objects with
masses of 0.01 -- 20 $\msun$ must be less than 0.1 (increasing,
however, to $\simlt$1 for 60 -- 300 $\msun$ objects),  limits
determined by line-to-continuum microlensing
effects in quasars (Dalcanton et al.~1997).  These limits still allow the
Milky Way Halo to consist largely of such objects.
For comparison, Galactic chemical enrichment arguments, assuming a standard
initial stellar mass function and standard stellar evolution, limit
the remnants of Population III stars to
$\Omega \leq 0.001$ for 
remnants in the range 4 -- 200 $\msun$
(Carr, Bond, \& Arnett 1984).  However, there are no strong limits on
black holes or non-topological soliton states arising from the early Universe 
or on relics from a non-standard very early generation of stars.

More recently, the microlensing surveys
are setting the best
limits on the fraction of dark objects in our own dark halo
over a wide range of masses.
The above collaborations search for MACHOs using the fact that 
when a compact dark object passes
in front of a source star in a nearby dwarf galaxy, the source star suffers
a temporary magnification due to gravitational microlensing \cite{pac86}.
The duration of this magnification is determined by a combination of
the lens distance, velocity, and mass, and
this degeneracy means
that the lens mass cannot be uniquely determined for an individual lensing
event unless other information is available.  However,
for a given halo model an average over the lens density and velocity
distributions can be made and an average duration can be estimated
\cite{griest91}
\begin{equation}
\that \approx 130\sqrt{m/\msun} \, \rm days,
\label{eqt_hat}
\end{equation}
where $\that$ is the time for the source to cross the Einstein ring diameter.

The 13 -- 17 events discovered by the MACHO collaboration in the Large
Magellanic Cloud (LMC) have durations
ranging from about 30 days to about 130 days (Alcock et al.\ 2000a), 
indicating lens masses
in the $0.1\msun$ to $1 \msun$ range.  However, the number of events found,
while larger than the expected background of 2--4 events from known
stellar populations, is significantly less than the 60--80 events
expected if the dark matter consisted entirely of objects in this mass range.
Thus a most likely MACHO halo fraction of 20\% was found, and
a 100\% MACHO halo was ruled out at 95\% c.l.
Earlier EROS \cite{eros-ccd,macho-eros-spike} and
MACHO \cite{macho-spike,macho-eros-spike}
searched for events with durations of less than 10 days and found
none, limiting the MACHO fraction of dark matter to less than 25\%
in the range from $10^{-7}\msun$ to $10^{-4}\msun$.
Finally, and most powerfully for the high mass end,
the EROS collaboration \cite{eros-2}
did a combined analysis of all their
microlensing surveys of the Magellanic Clouds, and
set a 95\% c.l. limit that objects in
the $10^{-7}\msun$ to $4 \msun$ range do not constitute 100\% of the dark halo,
and objects less than $1 \msun$ contribute less than 40\% of the
dark halo.

In this letter we improve on these limits in the high mass region by
performing a search for events with durations longer than 150 days.
We did not find any such events and therefore can improve the
upper limit to around $30 \msun$.

\section{Data}
\label{sec-data}
The data used in this analysis is precisely the data used in
Alcock et al.\ (2000a; hereafter A2000a),
to which we refer the reader for details.
In brief, this dataset includes 5.7 years of data on 11.9 million stars
in the LMC.  These comprise 21,570 images taken
over 30 $42' \times 42'$ fields in two filter bands, with the number of
exposures per field ranging from 180 to 1338.  The photometry of these objects
was arranged in lightcurves and searched for microlensing using the
analysis and statistics described in A2000a.  Event selection was performed
using those statistics, and in this paper we consider selection criteria
similar to the set A selection criteria described in A2000a,
which was the more conservative
of the two sets of selection criteria used there.  After removal
of variable stars and background supernovae
this set of selection criteria gives 13 microlensing events,
with fit durations between 34 days and 103 days.
This corresponds to physical durations between 42 days and 126 days when
a statistical correction for blending is made.
We note that one event (event 22)
was considered marginal in A2000a and excluded by hand from event
set A for reasons described in A2000a.
Further study of this event shows that the source
is extended and contains emission lines that are not characteristic
of stellar objects.  Event 22 seems likely to be a supernova of exceptionally
long duration or an AGN in
a galaxy at redshift $z=0.23$ and, therefore, is very unlikely to
be microlensing. 
Our redshift is based upon spectra with wavelength coverage 4340-9017 angstrom
obtained with the DBS spectrograph on the 2.3m telescope at Siding Springs
Observatory.
The exclusion of event 22 is relevant for this paper since this event
is the longest duration candidate microlensing event with $\that=230$ days.
See A2000a and Alcock et al.\ (2000b; hereafter A2000b)
for details of data taking, analysis, and event selection.

\section{New Analysis}
\label{sec-analysis}
The spirit of the current analysis is similar to that used in the
Alcock et al.\ (1996) search for planetary mass dark matter, but applied
to the long-duration end of our data rather than the short-duration end.
We create a simple set of selection criteria similar to selection criteria
set A, but tailored to find only events with durations longer than 150 days.
No such events are found, so any Galactic dark matter model that predicts
more than 3 microlensing events is ruled out at 95\% c.l.
Many sets of selection criteria were explored, but since our final result
at the high mass end depends very little on which set of cuts we use,
we choose simply selection criteria A from A2000a
with the additional constraint ``$\that > 150$ days".
Thus the complete set of selection criteria we use is described in Table 3
of A2000a.
This set of cuts gives no candidate microlensing events.
Using this set of cuts, we then calculate the complete photometric
efficiency as a function of input $\that$ with the method described
briefly in A2000a and in detail in
A2000b and Vandehei (2000).  This efficiency calculation
takes into account inefficiencies caused by bad weather,
seeing, telescope slips, etc., and includes a careful treatment of
blending.   Blending occurs when a single photometered object actually
consists of several underlying stars, only one of which is microlensed.
See A2000b for a detailed discussion.
The resulting efficiency is shown in Figure 1, along with the efficiency
for criteria A from A2000a.
This efficiency is then convolved with the predicted distribution
of microlensing durations from a halo model to find an expected number of
microlensing detections if the Galactic halo consisted 100\% of MACHOs.

For simplicity, we use model S from Alcock et al.\ (1997)
which is given by

\begin{equation}
\label{eq-stdhalo}
\rho_H(r) = \rho_0 { R_0^2 + a^2 \over r^2 + a^2 }
\end{equation}
where $\rho_H$ is the halo density,
$\rho_0 = 0.0079 \msun \, \pc^{-3} $ is the
local dark matter density, $r$ is Galactocentric radius,
$R_0 = 8.5 \kpc$ is the Galactocentric
radius of the Sun, and $a = 5$ kpc is the halo core radius.
With the standard thin disk, this model has a total rotation speed at
50 kpc of 200 km/s, with 190 km/s coming from the halo, giving
a total halo mass of $4 \ten{11}\msun$ out to 50 kpc.
We assume an isotropic Maxwellian distribution of velocities
with a 1-D rms velocity of $155 \kms$,
and assume a $\delta$-function MACHO mass function
of arbitrary mass $m$.

In Figure~2 we plot the resulting expected number
of events as a function of MACHO mass.  
The number of events is Poisson distributed. Therefore, when the number of
expected events is $\alpha$, the probability of detecting 0 events is
$exp(-\alpha$). For $\alpha=3$, one has $P({\rm 0 events}) = exp(-3)= 0.05$.
Thus any model that predicts more that 3 events is ruled
out at 95\% c.l.  We note that if a continuous range of masses is ruled out,
then any mass function containing only masses in the
ruled out range is also ruled out (Griest 1991).

Using the number of expected events from Figure 2, we easily derive
Figure 3, the exclusion plot for the new analysis, with the area
above the solid line being ruled out at 95\% c.l.
We see that objects with masses between
$0.3 \msun$ and $30.0 \msun$ cannot make up 100\% of the dark halo in this
model.  
Combining these limits with earlier limits \cite{macho-eros-spike},
that are stronger at
lower masses where the microlensing surveys have their peak sensitivity,
we see that objects with masses under $10 \msun$ cannot make up more
that 40\% of the Galactic dark matter in this model.
Note that since the microlensing experiments can only detect MACHO dark
matter, the fractional limit on the dark halo mass is strongly dependent on
the total amount of dark matter assumed to exist in the dark halo.
Since this is quite uncertain, a more model independent way to
state these results is that at 95\% c.l., less that
$4 \ten{11} \msun$ in compact objects with masses
less than $30 \msun$ can be present
in the Milky Way dark halo, and less than $1.6 \ten{11} \msun$ can be
present in compact objects with masses less than $10 \msun$.
Limits on halo fraction will
scale roughly with the total mass out to 50 kpc in a given halo model;
that is, a model with twice as much dark matter will have a limit
of around 50\% rather than 100\% at $30 \msun$.

\section{Discussion}

The limits given in this paper are  the strongest to date on
compact halo objects with masses above 1 $\msun$.
In particular, black holes or other dark compact objects
with masses less than $30 \msun$ cannot make
up the bulk of the dark matter.

We do note, however, that the present survey/analysis does not have
much sensitivity to objects with masses greater $30 \msun$.
There is a large background of slowly varying variable stars which
must be removed, and our main signal-to-noise cuts are not very good
at distinguishing these from microlensing.  Thus we rely primarily
on a long flat baseline, and on a direct cut on the fit event 
duration ($\that<600$ days) to remove this background.
Unfortunately,
these cuts also limit our ability to detect long duration
events coming from high mass lenses.  
We expect that analysis of the complete 8-year data set
will go some way towards solving this problem and we expect to
be able to push the current limit to higher masses,
or to present long duration microlensing
events when that dataset has been analyzed.

\section*{Acknowledgements}
We are very grateful for the skilled support given our project
by the technical staffs at the Mt.~Stromlo and CTIO Observatories,
and in particular we would like to thank Simon Chan,  Glen Thorpe,
Susannah Sabine, and Michael McDonald, for their valuable assistance in
obtaining the data.
Work performed at the University of California Lawrence
Livermore National Laboratory is supported by the
U.S. Department of Energy under contract No.\ W7405-Eng-48.
Work performed by the Center for Particle Astrophysics personnel
is supported in part by the Office of Science and Technology Centers of
NSF under cooperative agreement AST-8809616.
Work performed at MSSSO is supported by the Bilateral Science
and Technology Program of the Australian Department of Industry, Technology
and Regional Development.
DM is also supported by Fondecyt 1990440.  CWS thanks
the Packard Foundation for their generous support.  WJS is supported
by a PPARC Advanced Fellowship. CAN was supported in part by an NPSC
Fellowship.
ND and KG were supported in part by the DOE under grant DEF03-90-ER 40546.
TV was supported in part by an IGPP grant.

KG thanks Donald Lynden-Bell and Sasha Dolgov for asking the questions
that inspired this work
and the Aspen Center for Physics where some of it was performed.


%

\onecolumn 



\twocolumn

\begin{figure}
\epsscale{0.9}  
\plotone{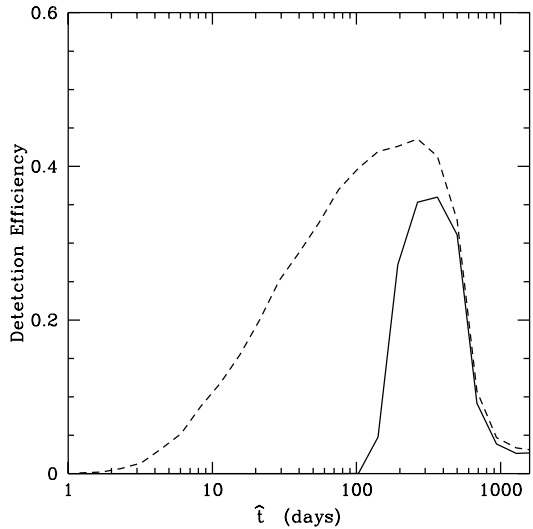}
\caption{ Microlensing detection efficiency 
for the 5.7-year MACHO data, as a function of event timescale $\that$.
The \textit{solid line} shows the photometric efficiency computed for cut
set A, with the additional constraint that $\that>150$ days.
The \textit{dashed line} (from A2000a) is the same but without the
additional constraint.
\label{fig-eff} }
\end{figure}

\begin{figure}
\epsscale{0.9}  
\plotone{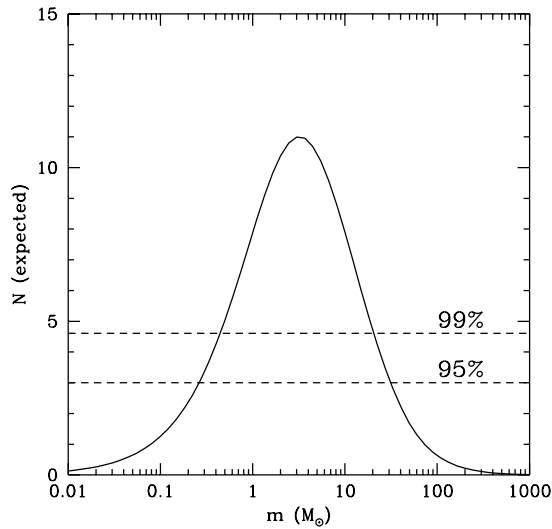}
\caption{Number of long duration events expected vs. lens mass for halo
model S.  The dashed lines drawn at $N=3$ and $N=4.6$ indicate 
the 95\% c.l. and the 99\% c.l. limit respectively.  Masses above these
lines are ruled out at their respective confidence limits.
\label{fig-nexp} }
\end{figure}

\begin{figure}
\epsscale{0.9}  
\plotone{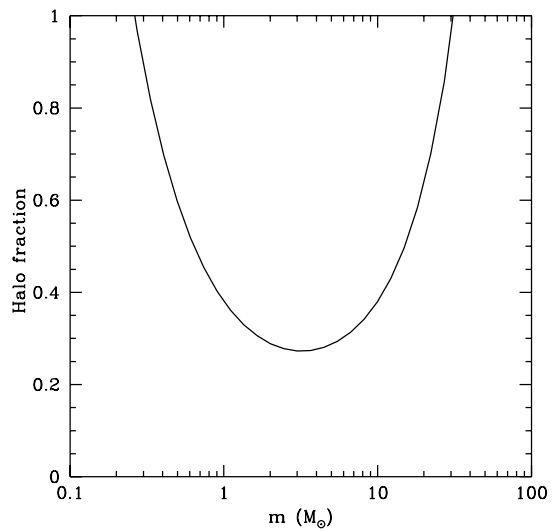}
\caption{Halo fraction upper limit as a function of lens
mass for model S.  The region above the line is ruled out at 95\% c.l.
This model contains $4 \ten{11} \msun$ within 50 kpc, so a less model
dependent result can be found by reading the ordinate as
``Halo mass in MACHOs/($4 \ten{11} \msun$)".
\label{fig-limit}}
\end{figure}

\end{document}